# Elastic deformation of soft coatings due to lubrication forces

Yumo Wang, Matthew R. Tan and Joelle Frechette

Elastic deformation of rigid materials with soft coatings (stratified materials) due to lubrication forces can also alter the interpretation of dynamic surface forces measurements and prevent contact formation between approaching surfaces. Understanding the role of elastic deformation on the process of fluid drainage is necessary, and the case where one (or both) of the interacting materials consists of a rigid substrate with a soft coating is still limited. We combine lubrication theory and solid linear elasticity to describe the dynamic of fluid drainage past a compliant stratified boundary. The analysis presented covers the full range of coating thicknesses, from an elastic foundation to a half-space for an incompressible coating. We decouple the individual contributions of the coating thickness and material properties on the elastic deformation, hydrodynamic forces, and fluid film thickness. We obtain a simple expression for the shift in contact position during force measurements that is valid for many experimental conditions. We compare directly the effect of stratification on the out-of-contact deformation to the well-known effect of stratification on indentation. We show that corrections developed for stratification in contact mechanics are not applicable to elastohydrodynamic deformation. Finally, we provide generalized contour maps that can be employed directly to estimate the elastic deformation present in most dynamic surface force measurements.

## Introduction

Rigid materials with soft coatings are ubiquitous in tribology,[1] microfluidic devices,[2] biomaterials,[3] or colloidal and particulate systems.[4] Under many practical settings they are employed in fluid environments where they are in close proximity to another surface. Under these conditions viscous forces due to the relative movement of the two surfaces can exert fluid pressures and cause elastic deformation (see Figure 1). More specifically, lubrication forces can lead to elastic deformation, also known as elastohydrodynamic deformation or EHD, which can prevent contact formation as fluid drains from a gap separating compliant materials.[5-7] If unaccounted for, EHD can lead to the misinterpretation of dynamic surface forces measurements. The surface forces apparatus (SFA)[6] or the atomic force microscope (AFM)[8] are commonly employed under dynamic conditions and can be used with soft materials. In particular, an exact description of lubrication forces is necessary to rely on dynamic surface forces measurements for the characterization of, for example, conservative surface forces,[9] fluid structure,[10] or surface slip[11] on compliant materials. In addition, recent reports show that ignoring the effect of elastic deformation can lead to a misinterpretation of the force data, contact position, and slip at the solid-liquid surfaces.[8, 12]

Most previous efforts to describe unsteady normal fluid drainage past an elastic boundary studied the case where the soft materials could be considered a half-space. For instance, Davis et al. studied normal elastohydrodynamic collisions between particles.[13] Kaveh et al.[7] modeled dynamic force measurement using AFM on thick PDMS coatings using Finite Element Method. These approximations are, however, invalid for stratified materials (here the case of a rigid material with a compliant coating). In particular, there are no solutions for drainage past elastic coatings of finite thicknesses: coatings that do not fall within the limiting cases of an elastic foundation or a half-space.[14-16] More specifically, the unsteady case where a surface initially at rest moves toward a static one at a constant drive velocity is of particular importance in the measurement of dynamic surface forces and has yet to be investigated.[17, 18] Finally, a cantilever spring needs to be incorporated to the model description to be applicable for surface forces measurements where the forces are measured via the deflection of a cantilever spring. Therefore, a treatment for unsteady fluid drainage in the presence of a stratified material is necessary to both interpret dynamic surface forces measurements and to engineer soft coatings that are subjected to viscous forces.

In this paper, we aim to answer the following questions. 1) How do the thickness and elasticity of a coating alter the drainage process? 2) How would elastic deformation affect dynamic surface forces measurements, and 3) Would corrections for stratification developed for indentation (contact mechanics) be applicable for EHD? To answer these questions, we develop a general solution for unsteady normal fluid drainage in the sphere-plane geometry that is valid for all coating thicknesses (including intermediate coating thicknesses) and that also include the effect of the force measuring spring. We non-dimensionalize the governing equations such that the effects of thickness and material property are naturally decoupled. As a result, the transition between the two known limiting cases for coating thicknesses can be fully visualized. Throughout, we discuss the coupling between surface deformability, coating thickness, and spring stiffness. We also provide numerical results for both rigid and soft springs, and derive a scaling for infinitely rigid spring that can be readily applied to many experimental settings.

*Department of Chemical and Biomolecular Engineering, Hopkins Extreme Materials Institute, Johns Hopkins University, Baltimore, MD, 21218, USA.*
*Email: jfrechette@jhu.edu*
† Electronic Supplementary Information (ESI) available: Details of numerical algorithm. Verification of present model to known limits.

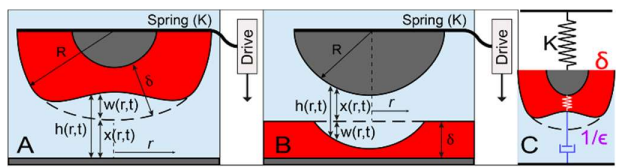

**Fig. 1.** Schematic illustrating the geometrical and experimental configuration investigated. (A, B): Mathematical equivalent configurations when R >> h. Dark grey: Rigid materials. Red: deformable material. Black bar: connecting spring. Definitions of variables: h(r,t): Separation between surfaces or fluid film thickness. w(r,t): Surface deformation. X(r,t): Undeformed separation. $\delta$ : thickness of elastic coating. $\varepsilon$ : Elasticity parameter (defined in Equation 9). Figures not plotted to scale (C) Mechanical element during fluid drainage in force-measuring scenario.

The paper is organized as follows: in the theoretical development section, we review prior theoretical treatments for elastohydrodynamic deformation and present the basic physical assumptions and the mathematical formulation governing our model. In the results and discussion section we first show the effects of film thickness for a fixed elasticity. We then expand our results to a wide range of elasticity and spring constants. We finally compare the effect of stratification on the force-deformation curves in elastohydrodynamic drainage to the ones that would be obtained from a spherical indenter.

## Theoretical development
### Background

Theoretical and experimental investigations of fluid drainage between stratified materials is relatively scarce in the literature. Balmforth et al.[19] developed a detailed analysis to characterize the sedimentation of spheres on different soft structures (foundation, half-space, beam, membrane) under gravity. Later, Leroy and Charlaix[20] developed a theoretical framework to characterize regimes where fluid viscosity or the film elasticity dominate the force response caused by an oscillatory probe in the limits of thin or thick elastic coatings (and steady-state). Leroy et al.[21] later verified their predictions using a modified SFA. More specifically, they showcased the sensitivity of the fluid drainage to the mechanical properties of the bounding surfaces by extracting the elastic modulus of the coatings from out-of-contact rheological experiments. The related problem of steady-state elastohydrodynamic lubrication (EHL) for lateral motion (sliding) has been studied[22] and an analysis valid for all film thicknesses and elasticity was developed by the Mahadevan group.[10, 23]

However, despite these recent advances, some challenges still remain in trying to describe unsteady normal fluid drainage bounded by stratified materials. Predictions for the unsteady case are not available. For instance, the effects of a coating's thickness and compliance are generally combined into a single dimensionless parameter,[19, 20] making difficult to extract the individual contribution originating from materials properties from those caused by geometry (film thickness). More importantly, treatment for cases where the stratified materials do not fall within a known limiting case such as an elastic foundation or a thick film (for example for films with thickness comparable to the hydrodynamic radius), have also largely been ignored.

### Problem formulation

Consider normal fluid drainage in the presence of a stratified elastic boundary in the sphere-plane geometry (see Fig. 1A). This geometry is commonly employed in colloidal probe AFM, and many other similar instruments designed to measure surface forces and adhesion.[24-27] This geometry is also equivalent to the cross-cylinder configuration employed in the Surface Forces Apparatus (SFA).[28, 29] As an analogy to the case of a rigid indenter on a soft wall, we set the spherical probe as rigid while the plate is stratified and consists of a rigid support coated with a more compliant material (Fig. 1B). However, the analysis presented here can easily be extended to any axisymmetric geometry and for the case where an elastic coating is present on both surfaces.[10]

The fluid flow is described by the lubrication approximation, which is valid in the limit of low Reynolds number and when the central fluid film thickness, $h(r=0,t)$, is small compared to the sphere radius ($R$), with $r$ being the radial coordinate and $t$ the time. If we assume the no-slip boundary condition on both surfaces and continuum fluid phase, the axisymmetric drainage and infusion of a Newtonian fluid (viscosity $\eta$) from a thin gap is given by:

$$\frac{\partial h(r,t)}{\partial t} = \frac{1}{12\eta r}\frac{\partial}{\partial r}\left(rh^3\frac{\partial p(r,t)}{\partial r}\right), \quad (1)$$

showing that the fluid pressure distribution, $p(r,t)$, between the two surfaces is related to the shape of the gap and to the rate at which the surfaces approach one another.

The fluid pressure distribution acts as a normal stress on the bounding stratified surface and causes elastic deformation. We use the linear elasticity theory for stratified materials to describe the mechanical response of the surface coating where the source of surface stress is the fluid pressure distribution. In our present geometry, the radius of the sphere is much greater than the fluid film thickness ($R \gg h$), so that the local curvature of spherical surface is very small. In this case, the normal stress $\sigma_N$ dominates over tangential stress $\sigma_T$ in the solid material, because $\sigma_T/\sigma_N \sim \sqrt{h/R} \ll 1$.[20] As a result, we can take the fluid pressure as the normal boundary pressure and neglect the shear stress at the surface. This is also the reason why the Fig. 1A and Fig. 1B are mathematically identical configurations (only when $R \gg h$): we inherently ignored the local curvature when applying the surface stress, so that the calculated asymmetrical deformation will not be affected by which surface it is on. Within the framework of linear elasticity, the deformation distribution of an elastic layer can be acquired by solving the biharmonic equation, which can be reduced to an ODE by using Hankel transforms in cylindrical coordinates. The four boundary conditions in our configuration are an axisymmetric normal stress and a negligible shear stress on top of soft coating, together with sticky boundary conditions on bottom surface. For this case, a closed form for the surface deformation, which is needed to calculate $h(r,t)$ in Equation 1 was derived previously in the context of indentation,[30, 31] and used by others in the context of elastohydrodynamics.[20, 32] The surface deformation can be calculated from:

$$w(r) = \int_0^\infty \frac{2}{E^*\xi} X(\xi\delta) Z(\xi) J_0(\xi r) d\xi \quad (2)$$

Where

$$X(\xi\delta) = \frac{\gamma(1-e^{-4\xi\delta}) - 4\xi\delta e^{-2\xi\delta}}{\gamma(1+e^{-4\xi\delta}) + (\gamma^2+1+4(\xi\delta)^2)e^{-2\xi\delta}}; \quad \gamma = 3-4\nu \quad (3)$$

And

$$Z(\xi) = \xi \int_0^\infty rp(r)J_0(\xi r)dr \qquad (4)$$

In Equations 2-4, $p(r,t)$ represent the applied boundary stress, which is the local liquid pressure on the surface at a given radial position $r$. $Z(\xi)$ is the modified Hankel transform of the pressure, in which $J_0(\xi r)$ is the 0$^{th}$-order Bessel function of the first kind. $\delta$ is the compliant film thickness (illustrated in Fig.1), $\nu$ is Poisson's ratio of the soft coating, $\xi$ is the Hankel transform variable, and $E^*$ is the reduced Young's modulus. Leroy and Charlaix used the approach outlined in Equations 2-4 to characterize the force response of an oscillatory motion on a thin or a thick coating mediated by a liquid.[20] Here we rely on Equations 2-4, but applied them into an unsteady continuous approach model to update the surface separation. We also verify the results by solving biharmonic equation numerically without using the Equation 2-4 for a few cases, using the method presented in Ref[33] and found the results to be identical. Throughout, the strain in the elastic coating always remains within the bounds of linear elasticity theory, the maximum normal strain we predict is ~6% and for the great majority of the reported results the strain is less than 2%, which falls safely in the linear regime, although the deformation of the coating can be of order of the fluid film thickness.[12]

During the measurement of dynamic surface forces, one surface is typically mounted on a cantilever spring. In this case, the surface separation is different from the displacement of the driving motor due to the deflection of the spring. For example, if the surfaces are being brought closer at a constant drive velocity $V$ (positive when approaching), a decrease in the thickness of the fluid gap leads to an increase in fluid pressure. As a result, the spring deflects (illustrated in Fig.1B) to maintain mechanical equilibrium, which leads to a deceleration of the relative movements between the two surfaces. In other words, deflection of the cantilever decreases the rate of fluid drainage in the gap. Another equation describing the mechanical coupling between the spring and hydrodynamic forces is therefore necessary to describe the drainage process. The spring and hydrodynamic forces can be evaluated at any radial positions in the gap, and for convenience we calculate it based on the central position value as shown in Equation 5:

$$F_s = kS(t) = k(h - h_0 + Vt - w) = F_H = \int_0^R 2\pi rp(r)dr \qquad (5)$$

In Equation 5, $F_s$ is the spring force which is balanced by the hydrodynamic force, $F_H$. The spring force is a product of the spring constant, $k$, and the deflection, $S(t)$, which is calculated from the initial separation at the centerpoint, $h_0 = h(0,0)$.

**Non-dimensionalization and discretization**

To simplify our calculation, we rescale the Equations 1-5 above to make them dimensionless (see supporting information). The resulting dimensionless lubrication equation is:

$$\frac{\partial \hat{h}}{\partial \hat{t}} = \frac{1}{12\hat{r}} \frac{\partial}{\partial \hat{r}}\left(\hat{r}\hat{h}^3 \frac{\partial \hat{p}}{\partial \hat{r}}\right) \qquad (6)$$

And dimensionless variables for the lubrication equation are:

$$\hat{r} = \frac{r}{\sqrt{Rh_0}}, \qquad \hat{h} = h/h_0,$$

$$\hat{t} = \frac{V}{h_0}t, \qquad \hat{p} = \frac{h_0^2}{\eta RV}p \qquad (7)$$

We can further non-dimensionalize the elastic deformation equation (2) to obtain:

$$\hat{w}(\hat{r}) = \varepsilon \int_0^\infty \frac{2}{\hat{\xi}} X(\hat{\xi}T)\hat{Z}(\hat{\xi})J_0(\hat{\xi}\hat{r})d\hat{\xi} \qquad (8)$$

With the following dimensionless variables:

$$\hat{\xi} = \xi\sqrt{Rh_0}, \qquad T = \delta/\sqrt{Rh_0},$$

$$\hat{Z} = \frac{1}{\eta V}\left(\frac{h_0}{R}\right)^{1.5} Z, \quad \hat{w} = \frac{w}{h_0}, \quad \varepsilon = \frac{\eta V R^{1.5}}{E^* h_0^{2.5}}, \qquad (9)$$

and the force balance (equation 5) becomes:

$$\hat{F}(\hat{t}) = K(\hat{h} - 1 + \hat{t} - \hat{w}) = K(\hat{x} - 1 + \hat{t}), \quad K = \frac{kh_0^2}{\eta R^2 V} \qquad (10)$$

Based on non-dimensionalization (Equations 6-10), we find that normal fluid drainage past a stratified material can be fully captured by three key dimensionless parameters and summarized in table 1. Note that all the non-dimensional variables are derived based on their governing equations, except for normal deformation, $\hat{w}$. Previous studies investigated fluid drainage for two limiting cases: very thin and thick elastic films.[19, 20] In these limiting cases, $X(\xi\delta)$ can be expanded, so that the terms containing Poisson's ratio and terms containing $\xi$ can be separated, allowing for the terms containing Poisson's ratio in normal deformation to be extracted directly from the integration, and form dimensionless parameters for deformation.

Unfortunately, relying on a Taylor expansion for the limiting cases leads to different scaling parameter for each case, making it difficult to compare the same coating material but with varying thicknesses. More importantly, this method also prevents studying films that do not fall into these limiting cases. An option is to set the Poisson's ratio to a constant value for the equations to be made dimensionless due to the non-linear dependence of $X(\xi\delta)$ with respect to the Poisson's ratio in equation 3. This is the avenue we decided to pursue here as we seek a general framework that is valid for all thicknesses and elasticity, including intermediate film thicknesses. Therefore, we did not expand $X(\xi\delta)$ as done in previous work,[19, 20] but instead, we defined an elasticity parameter, $\varepsilon$, based on reduced Young's modulus, to non-dimensionalize the normal deformation $w$ and set the Poisson's ratio to a constant value of $\nu = 0.5$, to cover most polymeric materials and rubbers.[34] Note that the Poisson's ratio is also set at 0.5 in $X(\xi\delta)$ (Equation 3). Thus, the results presented here would change for different Poisson's ratio, and the effect of compressibility will be the subject of a future study. The fact that the effect of compressibility cannot be isolated is inherently due to the varying effects of compressibility over film thickness $\delta$ in determining the normal deformation in the presence of a sticky rigid substrate boundary condition, especially for thinner coatings.

Table 1 Definition of the three key dimensionless parameters

| Parameter | Relationship | Interpretation |
|---|---|---|
| Spring | $K = \dfrac{kh_0^2}{\eta R^2 V}$ (10) | $\dfrac{spring\ forces}{viscous\ forces}$ |
| Elasticity | $\varepsilon = \dfrac{\eta V R^{1.5}}{E^* h_0^{2.5}}$ (9) | $\dfrac{compliance\ of\ coating}{viscous\ forces}$ |
| Thickness | $T = \dfrac{\delta}{\sqrt{Rh_0}}$ (9) | $\dfrac{thickness\ of\ elastic\ coating}{hydrodynamic\ radius}$ |

We solve Equations 6, 8 and 10 simultaneously to obtain the spatiotemporal surface deformation, viscous forces, fluid pressure, and fluid film thickness for coating of varying thicknesses and elasticity. The equations are solved numerically, and the details of numerical algorithm is available (Fig.S1, ESI†). We neglect conservative surface forces, such as van der Waals interactions, because under most circumstances of interest here the hydrodynamic interactions dominate over surface forces.[35, 36] In particular, deformation of compliant surfaces increases the separation between the two surfaces, which diminishes the contribution of surface forces. However, surface forces could become very important when the two surfaces are close to contact, especially when the motor is stopped suddenly, allowing the spring to relax and the surfaces to make contact. In that case, the additional pressure from van der Waals and Double layer interactions can be incorporated into boundary stress of elastic layer.[37] We start the calculations from the initial separation at $\hat{t}=0$ when the motor starts moving (from rest) at a constant drive velocity, until cutoff time, when the dimensional fluid film thickness is $h$ = 10 nm (or dimensionless $h$ = 0.004) to avoid singularity at $h$ = 0. We only consider the central part of the spherical surface (for $r<0.1R$) where the drainage pressure is the largest, and the pressure past $r=0.1R$ is set to be 0 because it is negligible compare to that of the center. The central region ($r<0.1R$) is discretized into 500 evenly spaced elements. We perform a convergence test for all discretized variables, including radial position and time. The mesh size on both $\hat{t}$ and $\hat{r}$ are decreased until the change in the calculated hydrodynamic force between subsequent iterations is less than 1%. In all of our results, the increment in dimensionless time is set as 0.00667, and the mesh size in dimensionless radial position is 0.0163. Note that the mesh size here not only affects the convergence results, but also has an effect on integration of Hankel transform variables (Equations 2-4). A fairly coarse mesh in the r-direction would cause the integration function $X(\xi\delta)$ to fluctuate over spatial variable $\xi$ which results in errors in deformation calculation. We validate our results by comparing it to two known limits: the Reynolds' theory for rigid surfaces[38] and DSH (Davis-Seyrassol-Hinch) model between elastic half-space.[13] (Fig. S2, ESI†)

## Results & discussions

### Choice and relevance of parameters investigated

The effects of our three key dimensionless parameters (Spring $K$, Elasticity $\varepsilon$, and Thickness $T$) on the drainage dynamic are evaluated systematically in the following sections. Throughout, we highlight combinations of parameters that are relevant to experimental conditions encountered in dynamic surface forces measurements, microfluidics, and adhesion. We first analyses the effect of the coating thickness for thin ($T<0.1$), intermediate ($T \sim 0.1-1$), and thick films ($T>1$), while keeping the other two parameters constant. We then expand to study the role of elasticity by varying $\varepsilon$ from $\varepsilon=10^{-2}$ at which the coating is very deformable, to $\varepsilon=10^{-4}$ at which the deformation due to drainage forces is hardly measurable. Finally we introduce the effect of the spring parameter and study $K=10$, $K=200$ and $K \sim \infty$, to cover most common experimental configurations ($R$, $\eta$, $V$, $k$), encountered in surface forces measurements.

We show in Figure 2 combinations of $K$ and $\varepsilon$ from prior work reported in the literature.[7, 8, 12, 39] For each experimental configuration ($R$, $\eta$, $V$, $k$) the shaded areas show how changing the modulus of the coating, from 0.5 MPa to 10 MPa (elastomers), would change $K$ and $\varepsilon$. Lowering the elasticity further would bring most experimental conditions into the $K \sim \infty$ regime, which is solved here. In dynamic force measurements, the value of initial separation, $h_o$, can be easily controlled. All three key parameters are a function of $h_o$, and as a result $K \sim (1/\varepsilon)^{0.8}$. Therefore, changing $h_o$ in an experiment moves the parameters on a line with a slope of 0.8 in Fig.2. For most parameter sets in highlighted area, the numerical predictions of EHD experiments can be either directly obtained from the results presented here (dashed lines and green regime), or easily converted to the discussed regime (dashed lines and green regime) by simply shifting $h_o$ in experiments. Additionally, the numerical values for the three parameters in table 1 can be manipulated individually (without affecting the other two) by changing spring constant, thickness of the coating, and the reduced modulus $E^*$ of the coating to reach a specific case of $K$ vs. $1/\varepsilon$. If $h_o$ is set to be a specific value and experimental parameters do not fall exactly in the discussed regime, an equivalent result can be obtained with a different $k$ (or $E^*$) but with identical other parameters as a point of comparison. An additional limiting case not investigated here would be for a surface connected to a cantilever with a spring constant of zero. In that case the motor motion does not cause displacement or deformation, and reverts to the scenario of particle collisions with an approach at an initial velocity and a zero net-force collision.[13]

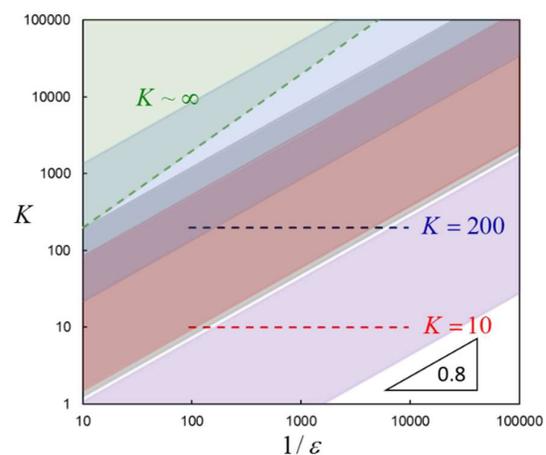

**Fig. 2.** Our choice of parameters for $K$ and $\varepsilon$ plotted according to common experimental conditions. The three dashed lines indicate the range of spring parameter and elasticity investigated here. The shaded areas represent experimental configuration of prior experimental studies for reduced modulus of the soft coating between 0.5-10 MPa, where purple: Charrault et al.[8], blue: Kaveh et al.[7] red: Wang et al.[12], and grey: Honig et al.[39] A slope of 0.8 in log-log scale is illustrated in the bottom right corner. The green regime above the green dashed line indicates $K \sim \infty$ which is plotted as $K > 20(1/\varepsilon)$ here.

**Effect of film thickness**

We first investigate how the thickness of the deformable film influences fluid drainage. We set the elasticity parameter at $\varepsilon = 0.01$ to fall in a regime where the elastic deformation is limited by the thickness and not the elasticity of the coating, and the spring parameter at $K = 200$. This choice of $K$ maps many typical experimental conditions (see Fig.2), The limit of $K = 200$ represent the condition in which the spring deflection need to be considered in determining the undeformed position $x(r,t)$, but does not have a strong effect on the drainage process.

We consider the approach of a rigid indenter toward a stratified wall at a constant drive velocity (Fig. 1B). We track the central separation, $\hat{h}(0,t)$, for fluid drainage past coatings of varying thicknesses (Fig.3A). Two established limiting cases are also shown in Figure 3: calculations using the DSH half-space model[13] in grey solid lines and Reynolds theory for rigid surfaces[38] in dashed black lines. We see that as the coating thickness increases the drainage dynamics transitions from the rigid to the half-space limits. We also see that our treatment for stratified materials recovers these two limits without having to modify the definition of the dimensionless parameters. The thickest coating, which is plotted in red lines (T = 82), overlaps with the half-space grey line. Similarly, the thinnest coating (green line, T = 0.02) overlaps the Reynolds' limit. Note that the motor drive position (the straight brown line) does not overlap with Reynolds' theory (black dashed line) in Fig. 3A due to the spring deflection.

We see that the approach slows down because elastic deformation competes with fluid drainage (decrease in the slope in Fig.3A). In fact, the approach essentially reaches a halt and a finite, and often significant, fluid film is trapped between the surfaces even if the motor keeps pushing the surfaces towards each other (motor position decreases linearly with time). The difference between the curve corresponding to the drainage past a rigid substrate and any curve for a coating of thickness T is the (dimensionless) normal deformation of the coating at the centerpoint (shown in Fig.3B). Due to the constraints imposed by the underlying substrate, decreasing the coating thickness leads to smaller normal deformation at the centerpoint than for thicker films and, as a result, a thinner fluid film trapped between the surfaces. We see that beyond $\hat{t} = 1$, the deformation for thick film (T = 82) approaches a slope of 1, meaning that most of the motor displacement goes into elastic deformation of the coating and not towards fluid drainage (Fig.3B).

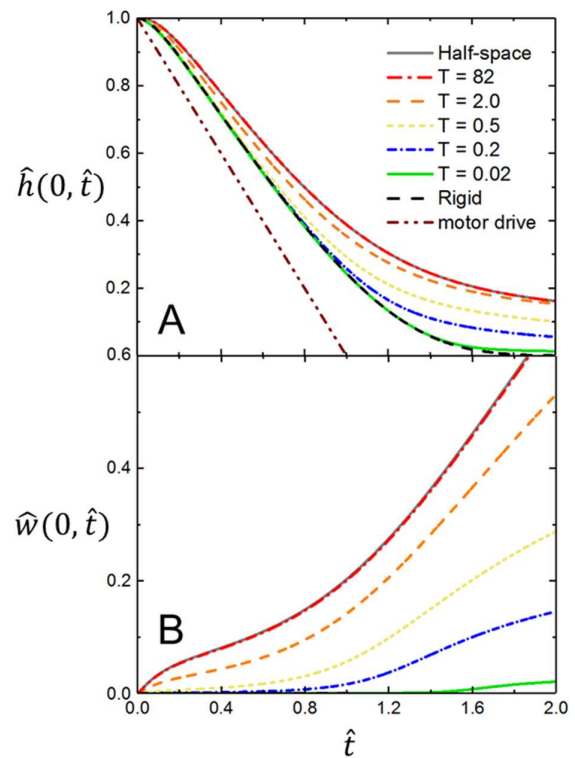

**Fig. 3.** (A) Central separation and (B) central deformation during the approach of a rigid indenter towards a stratified material as illustrated in Fig.1B. The coating has an elasticity parameter $\varepsilon = 0.01$ and a spring parameter of $K = 200$. In (A) the dashed black line indicates Reynolds' theory for rigid materials, an equivalence of thickness $T = 0$, and dashed double dots line in brown represent the motor drive position. In A-B the grey lines represent the DSH theory for elastic half-space, an equivalence of $T = R$. All other thicknesses are indicated in the inset of (A).

**Predicted force measurements**

Stratification would also have a strong effect on the hydrodynamic forces measured with the SFA or AFM. In Fig.4 we show typical force curves measured by a cantilever spring ($K = 200$). To mirror experiments with the SFA the forces are plotted with respect to the absolute separation (Fig.4A), while the forces are plotted with respect to the undeformed separation $\hat{x}(0,\hat{t})$ (Fig.4B) to represent more closely experiments performed with the AFM. Initially, even if deformation is present the hydrodynamic forces are nearly the same and independent of the thickness of the elastic coating. However, from $\hat{t} \approx 0.5$ (or $\hat{h}(0,\hat{t}) \approx 0.3$) the role played by the coating thickness becomes more noticeable. In particular, the fact that a finite fluid film is trapped between the surfaces at long times, see Fig.3, leads to a hydrodynamic force that essentially acts as a hard wall when plotted with respect to the central fluid film thickness (Fig.4A). As predicted, the force on thicker films deviate from Reynolds' earlier compared to thinner coatings, because a thicker elastic layer is less constrained by the underlying rigid substrate leading to larger deformation. Therefore, the assumption of undeformable surfaces in

Reynolds' theory start to fail at larger separation compared to the case of thinner coating.

The role played by elastic deformation on hydrodynamic repulsion depends on the comparison point. At a given undeformed separation, increasing the coating thickness leads to a reduction of the hydrodynamic forces (Fig.4B). On the other hand, the force in general increases with the coating thickness at a given central separation (Fig.4A), except for very large

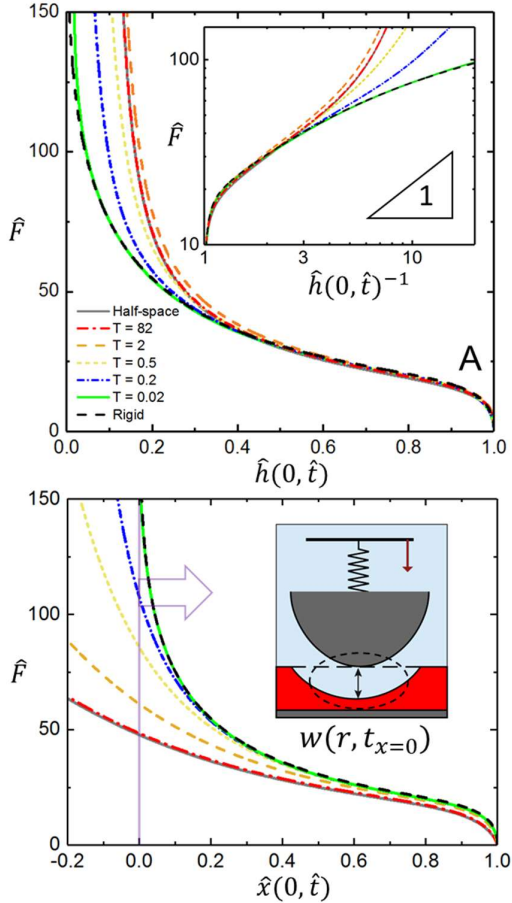

**Fig. 4.** Hydrodynamic forces during the approach of a rigid indentor towards a stratified material as a function of (A) central separation, (B) undeformed central separation, (inset of A) inverse of the central separation. The coating has an elasticity parameter of $\varepsilon = 0.01$ and a spring parameter of $K = 200$. The color schemes are the same as in Figure 3. The black triangle in inset of A indicates the slope of 1. The purple line in B indicate the $\hat{x}=0$ location. The inset of B shows the configuration at $x=0$ and define offset contact deformation $w(r,t_{x=0})$.

thickness, which is because of the initialization of experiments and is discussed in ESI in more detail. We discussed previously this apparent "contradiction" for the case of an of elastic half-space[12] and generalize it here for all coating thicknesses. Normal elastic deformation is always larger for thicker films. Therefore, at a given time or undeformed separation the larger deformation observed with a thicker coating leads to a reduction in the central fluid pressure in the lubrication equation, and a lower total integrated force. In contrast, at a given central separation, the larger deformation in thicker films leads to a flatter surface profile, and as a result to a broader pressure distribution. Therefore, an increase in the hydrodynamic force is observed in thicker films when compared at a given central separation (Fig. 4A). For measurements using techniques such as the AFM, the hydrodynamic force is obtained with respect to the undeformed separation due to the positioning of the photodetector. As a result, dynamic AFM measurements would show a decrease in repulsion with an increase in film thickness.[7, 8] On the other hand, in the SFA or with other force measurement techniques based on interferometry or profilometry, the raw data is the surface separation, h(t), and an increase in hydrodynamic repulsion would be observed for thicker elastic coatings. [6, 12]

Finally, the Reynolds' equation predicts that the repulsive force should scale with $1/\hat{h}(0,\hat{t})$ for rigid surfaces moving at a constant velocity, we highlight the deviation from that relationship in inset of Fig.4A. For rigid surface (black dashed line), the force scale with $1/\hat{h}(0,\hat{t})$ initially but lay under the slop of 1 as the separation become smaller. This is due to the fact that the spring decelerate the surface movement ($dh/dt < V$). However, as the coating thickness become larger, the increase in hydrodynamic force caused by deformation compensate for the spring effects, and the forces become stronger than predicted from Reynolds' equation. Note that deformation (with or without stratification) could easily mask slip at the solid-liquid interface, as both effects have a similar (but opposite) effect on the drainage process. This concern was raised previously by Vinogradova et al. [11, 40] The curve in the inset of Fig.4A can be employed to estimate the relative importance of coating thickness compared to slip in dynamic force measurements.

## Effects of elasticity

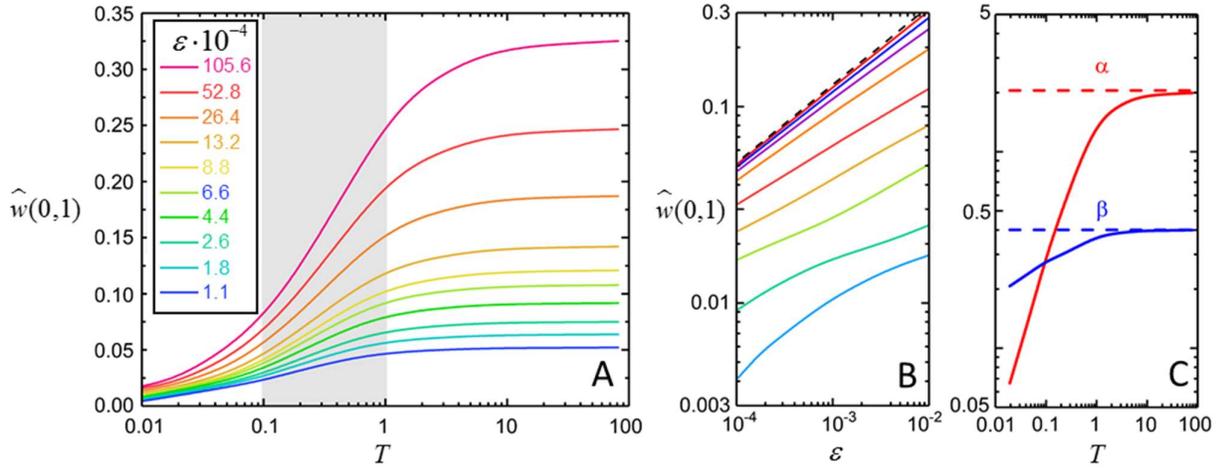

**Fig. 5:** Role of the thickness parameter on the offset contact deformation (fluid film thickness at $\hat{x}(0,\hat{t})=0$). (A) transition between an elastic foundation and a half-space for various elasticity (highlighted in the shaded area where T=0.1-1). (B) Offset contact deformation as a function of Elasticity for various thicknesses. From bottom to top: T = 0.01, 0.02, 0.05, 0.1, 0.2, 0.5, 1.0, 2.0, and 5.0. The half-space limit is indicated by the dashed black line. (C) Empirical parameter α and β used in Equation 12 as a function coating thickness. The dashed lines indicate the value of 2.05 and 0.4, which are expected for the half-space limit.

We introduce the "offset contact deformation" or $\widehat{w}(0,\hat{t}_{x=0})$, which is the deformation at the centerpoint when contact would occur in the absence of elastic deformation (illustrated in the inset of Fig.4B). The deformation at $\hat{x}(0,\hat{t})=0$ is important because in AFM experiments a strategy to define the contact position is to approach slowly (nearly quasi-statically) to minimize viscous forces and then record the position of the onset of the constant compliance regime. This position is then used as a reference contact position ($\hat{x}=0$) in dynamic experiments. The offset contact deformation defined here is therefore the deformation caused by viscous forces at this reference position. Knowledge of the deformation at $x=0$, which is $\hat{h}(0,1)=\widehat{w}(0,1)$ when $K\sim\infty$, is needed to shift the x-axis in a force-displacement curve to account for deformation.[7,8] We rely on the offset contact deformation to expand our investigation to the combined roles of the elasticity and thickness parameters in the limit of $K\sim\infty$. The effects of finite spring parameter will be discussed in the next section.

For any given elasticity parameter we observe three distinct regimes in the offset contact deformation as the thickness parameter increases (Fig.5A). For thin films ($T<0.1$, approximated from Fig. 5A) the offset contact deformation is small and depends more strongly on the thickness than on the elasticity of the coating. In this limit the stratification, i.e. the constraints imposed by the underlying rigid substrate, dictate the mechanical response over the material properties. For thick films ($T>1$) we find that the offset contact deformation depends is almost only a function of the elasticity parameter. This regime is the half-space limit where stratification does not have a significant effect on the mechanical response of the coating. This limit of thick films has been studied previously with the SFA and AFM.[7,12] For intermediate film thicknesses ($0.1<T<1$, highlighted grey region in Fig. 5A) both the mechanical properties of the coating (elasticity parameter) and stratification (thickness) determine the offset contact deformation. In this transition region, the offset contact deformation increases rapidly and is non-linear.

For almost all values of the thickness parameter, the offset contact deformation scales linearly with elasticity in a log-log plot (Fig.5B). The different regimes observed for an increase in thickness are not observed when increasing the elasticity. If we keep increasing the coating thickness (Fig. 5A) we will ultimately reach the half-space limit, in which the deformation is independent of the thickness. In contrast, an increase in elasticity will always results in increasing normal deformation. Therefore, the roles of the coating thickness and elasticity parameters are not interchangeable, and the interplay between the thickness and elasticity parameters can also be visualized by comparing Fig.5A-B.

The linear dependence in $\widehat{w}(0,\hat{t}_{x=0})$ with an increase in elasticity (Fig.5B) can be explained as follows. Consider a spherical elastic half-space subject to a uniformly distributed force on a projected area of radius $\sqrt{Rh}$, where the effective stiffness for the half-space can be approximated as $\pi E^*\sqrt{Rh}$.[41] When the force acting on the area is the Reynolds force: $6\pi\eta VR^2/h$, the central deformation can be then estimated as:

$$w(0,t) \approx \frac{6\pi\eta VR^2}{h} \cdot \frac{1}{\pi E^*\sqrt{Rh}} \quad . \quad (11)$$

At the point where $x(0,t)=0$, $h(0,t)=w(0,t)$. Therefore, in terms of dimensionless parameters, we obtain:

$$\widehat{w}(0,1) \approx (6\varepsilon)^{0.4} \quad (12)$$

a relationship plotted in Fig.5B as the black dashed line. As it is derived based on a half-space, we see that this approximation is recovered for thick films. For thinner films, despite the linear behavior down to T = 0.05, the slope and intercept are no longer 0.4 and $6^{0.4} \approx 2.05$, because the constraint from the substrate makes the effective stiffness of the surface no longer $\pi E^*\sqrt{Rh}$.

We obtain a simple and general expression for the offset contact deformation from the numerical results shown in Fig. 5B based on the relationship:

$$\hat{w}(0,1) \approx \alpha \varepsilon^\beta, \qquad (13)$$

where β is the slope and α is the intercept in Fig.5B. Eqn. 13 has an analogous form as Eqn. 12 but the slope and intercept were determined numerically. The values for $\alpha$ and $\beta$ are given as a function of film thickness in Fig.5C.

A quick and easy analytical determination of the shift in the contact position can be obtained from Eqn. 13, and is valid for a broad range of experimentally relevant conditions. Using the curve of Fig.5C one can directly estimate $\hat{w}(0,1)$ for any given $\varepsilon$ and T, where $\alpha$ and $\beta$ serve as a correction factors to account for the finite coating thickness. Note that this semi-empirical relationship works only when the spring constant $k$ is significantly larger than $\pi E^*\sqrt{Rh}$. If the spring stiffness is comparable to, or smaller than effective stiffness of surface, the deformation of the surface needs to be coupled with the spring deflection. When the spring constant is comparable to the elasticity, the limit of $K \sim \infty$ will give an upper bound for the shift in contact position, although for many experiments such as probe-tack tests[42], extensional rheology measurements,[43] and more generally measurements with stiff cantilevers and load cells, the $K \sim \infty$ situation are often encountered.

**Effects of spring parameter**

We can use simple scaling arguments to determine the relative contribution of the spring deflection to the drainage process. The cantilever shares the same hydrodynamic force as the deformable coating when it is mounted directly to one of the surface, as is the case for most surface forces measurements (Fig.1C).[44-46] If we model the experimental configuration shown in Fig.1C as a spring-dashpot and assume that the soft coating has an effective stiffness of $\pi E^*\sqrt{Rh}$ (half-space), then we can combine the two springs in series (cantilever spring and the deformable surface) into an effective spring with a stiffness of $k_{eff}$. The modified effective stiffness of the combined spring is $k_{eff} = \left(k^{-1} + (\pi E^*\sqrt{Rh})^{-1}\right)^{-1}$, which in dimensionless form returns to $\left(1/K + (\varepsilon/\hat{h}(0,\hat{t})^{0.5})\right)^{-1}$. Therefore, in the limit of $k \ll \pi E^*\sqrt{Rh}$, the decrease in surface separation during drainage will be dominated by the spring deflection and not by the properties of the coating. In this limit, the drainage process ($\hat{h}(\hat{t})$ and $\hat{F}(\hat{t})$, or $\hat{F}(\hat{h})$) can be simplified to the case of an infinitely rigid spring connected to a soft coating with effective stiffness $k_{eff}$ and is described by the predictions for $K \sim \infty$ shown in the previous section. This simple relation for the effective spring stiffness demonstrates that the spring contribution inherently varies during the drainage process. Initially, when $\hat{h}(0,\hat{t})$ is close to unity, the spring parameter $K$ should be compared with $1/\varepsilon$ in determining the effective spring constant of the system. As the fluid drains from the gap, $\hat{h}(0,\hat{t})$ decreases from one to zero and the surface elasticity plays an increasingly important role in the effective stiffness.

Finally, at long times $\hat{h}(0,\hat{t})$ will reach its asymptotic value (Fig.3A), making the spring contribution in the total stiffness of system independent of time. The coupling between the surface elasticity and the cantilever spring implies that the hydrodynamic forces will increase more rapidly with time for a stiffer cantilever spring. This trend is observed in Fig.6 where we plot the force vs time curves for different $K$ for a material with $\varepsilon = 0.0026$. The solid lines are for T = 0.2 and the dashed line is for T = 20, which is close to half-space. For stratified material, however, the effective stiffness of surface is larger than $\pi E^*\sqrt{Rh}$. According to this analysis, the contribution of the spring is more important compared to a half-space made of same material. An estimation on effective stiffness of soft coatings can be calculated from theories in contact mechanics, which we will discuss in later.

At long time, the force required for additional normal indentation (to decrease $\hat{x}$) increases (Fig.4B) and the change in the undeformed separation decreases over time. When $d\hat{x}(t)/dt \ll 1$, or when the change in undeformed separation is no longer comparable to the motor velocity, the force will increase linearly with time with a slope of $K$. This linear behavior is predicted by Equation 10, showing that the hydrodynamic force will increase linearly with a slope of K if $\hat{x}$ no longer decreases with time. As seen in Fig.6, all the force curves recover a linear regime with a slope that is exactly equal to the K value (as shown by triangle in the inset of Fig.6) at long times. This linear regime occurs for all film thicknesses, but it kicks in later for a thicker coating because there are no constraints from the substrate limiting the normal deformation (solid vs dashed green lines in Fig. 6). Once in the linear regime, any additional movement of the motor is transmitted to spring deflection, and therefore the motion of the motor has little effect of the fluid film thickness. This linear regime at long time is analogous to the "lock in regime" described by Charlaix et al., where the fluid cannot drain out because all surface displacement is transmitted to elastic deformation, not towards decreasing the fluid film thickness.[20] Our results here show that the smaller the spring constant is, a longer time is needed to enter this "lock in regime", although the repulsive force would be smaller compared to the one measured with a stiffer spring at the same time.

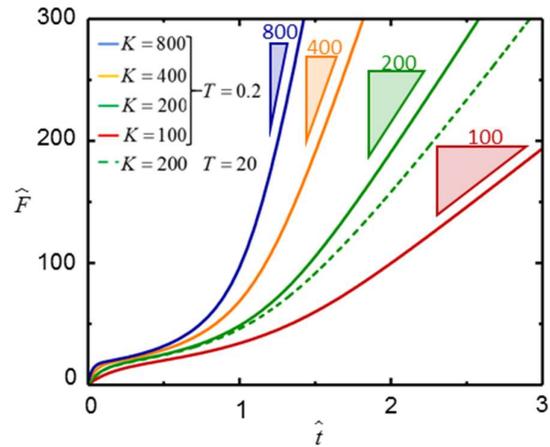

**Fig. 6.** Effect on the cantilever spring constant on the repulsive hydrodynamic force during approach. Dashed line: T = 20, solid lines: T = 0.2. Triangles: Slope of spring constant indicated by the legend. All variables are dimensionless.

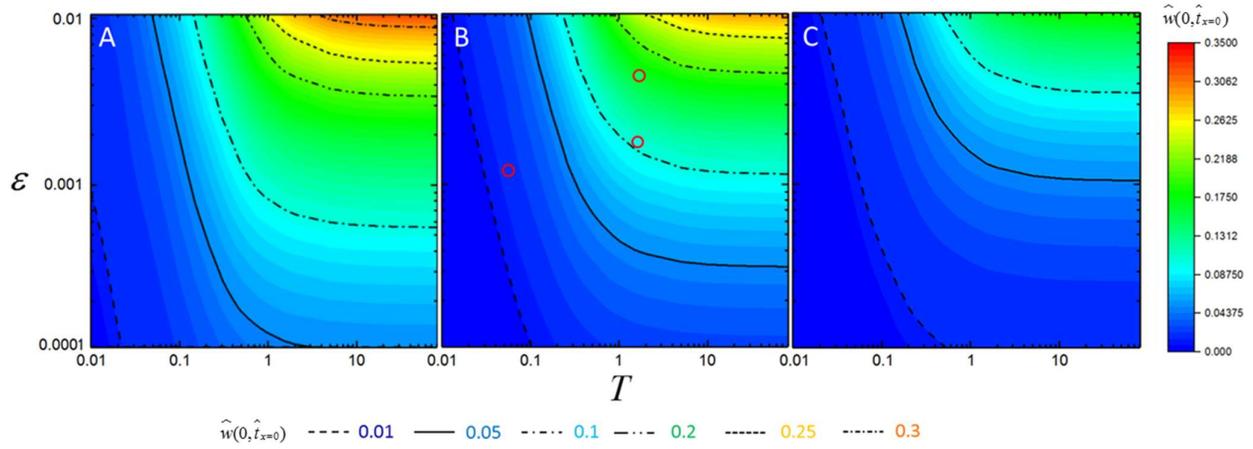

**Fig. 7.** Contour maps for the offset contact gap, $\hat{w}(0,\hat{t}_{x=0})$, defined as surface central separation at $\hat{x}=0$, across a wide range of elasticity and thickness parameters. The color gradient shows increasing values of $\hat{w}(0,\hat{t}_{x=0})$, indicated by the legend on the right. The lines show some selected values. The contour maps are shown for three spring parameters: (A) $K \sim \infty$ (B) $K=200$ and (C) $K=10$. The red circles in B indicate results retrieved from previous work.[12] The colors in the red circles indicate the experimentally measured values Which is compared with outer colors showing current model predictions.

For a finite spring constant, we do not have a simple analytical expression such as Eq. 13 to determine the offset contact deformation. Instead we created contour plots showing the offset contact deformation a function of elasticity and thickness for three spring parameters ($K \sim \infty$, $K=200$ and $K=10$), see Fig.7. Note that $\hat{w}(0,\hat{t}_{x=0})$ defined here is not necessarily $\hat{h}(0,1)$ because the deflection of the spring has an effect on contact time for a rigid surface. The different color in Fig.7 represent different magnitude of deformation. As expected, $\hat{w}(0,\hat{t}_{x=0})$ increases with both increasing elasticity and thickness, and reaches a plateau for thick films as it reaches the limiting case of a half-space. We find that the surface deformation is reduced significantly when the spring is more compliant (Fig.7B-C). That said, for a spring as soft as K=200, the offset contact deformation is still on the same order of magnitude as the one predicted from K = infinity. In fact, the larger the deformation is, the smaller the error is (for example, 9% error at the top-right corner, 95% error at bottom-left corner, others lay between these two limits). So the approximation based on K=infinity (which is analytical) is particularly helpful.

When $\hat{w}<0.05$, an analytical solution is available for fluid drainage.[13] For even smaller deformation, for example $\hat{w}<0.01$, the deformation can be safely neglected. The contour plots in Fig.7 can be used to determine the shift in contact position in experiments or to estimate the deformation should be incorporated in the analysis of surface forces data. Note that for urfaces on the same contour lines, the central deformations are the same, however the full deformation profile might be different. For Fig.7A ($K \sim \infty$) the plot can be fully approximated using results from Fig.5 and equation 13. To compare between experiments with different materials, a set of equivalent film thicknesses and spring constants can be found in the contor map to match the same offset central deformation. Experimental data from previous work[12] are plotted in Fig.7B in red circles. The color inside the circle indicates the measured offset central deformation. From the comparison between inner and outer color of red circles, we can see that the model works very well for determining the deformation of finite coatings, even though the spring parameter K are not precisely 200 for these three points (150-500).

**Comparison with contact mechanics approximations**

We investigate if treatments for the effect of stratification developed for indentation could be applied for elastohydrodynamic deformation. The effect of stratification on indentation has been recognized and studied in depth by Barthel et al[47, 48] and many others.[30, 31, 49-51] For the contact mechanics treatment, corrections for the penetration of indenter are available where the stratification effects are fully captured by an effective Young's modulus that is a function of the ratio between the contact radius and the coating layer thickness. With these corrections, the rest of the contact mechanics analysis follows that of a half-space. However, unlike the indentation case, the pressure distribution exerted during fluid drainage causes fluid flow and elastic deformation. This dual role makes the applicability of the contact mechanics approach questionable for continuous fluid drainage.

Consider the indentation with a spherical probe of a stratified material. The slope of the load-indentation depth is 1.5 in the limit of a half-space (based on Hertzian contact). For stratified materials, the slopes and intercepts increases because of the stiffening of rigid substrate. Corrections for stratification based on an "effective modulus" can be determined from the intercept of the load-indentation depth curves (log-log).[51]

We can compare directly indentation and EHD by plotting the load vs central deformation for different coating thicknesses for the two cases (Fig.8). We use our model to calculate the force-central deformation relationship for EHD for three different coating thicknesses as well as for the limit of a half-space, see the dashed lines in Fig.8 (for a fixed elasticity and spring parameter). For indentation (solid lines in Fig.8) the relationship for the applied force-central deformation is calculated from previous work for the same elasticity and film thicknesses as in the EHD case.[49] The values of the force and deformation are rendered dimensionless also using same parameters as for the EHD case.

Note that in Fig.8 the solid lines for indentation already account for stratification, and its effects are clearly visible. For example, the solid line with a thinner coating lies above the thicker coatings, indicating the stiffening effects of the supporting rigid substrate. Also, as the central deformation gets larger, the slopes of the finite thickness lines (blue, yellow, and red solid lines) are increasing slightly, indicating the increasingly important role of the supporting substrate as the indenter penetrates deeper. As a result, the effective modulus of the coating increases with an increase in the central deformation.

It is clear that for EHD the force-central deformation relationship is completely different than for indentation (Fig. 8). For a given force, the central deformation is always much lower for EHD than for indentation because energy is also dissipated via fluid drainage and not only via elastic deformation. Also, in contrast to the indentation case, we also see that the force-deformation relationship for EHD is highly non-linear (on a log-log plot). In fact, three regimes are clearly visible and highlighted by the shaded regions. Initially, the force increases linearly with a slope of 1 with deformation. This initial regime is caused by the spring deflection during start-up. When the motor starts to move, the deflection (and force) on the spring builds up and $\hat{h}$ remains mostly unchanged at 1 (visible during start up in Fig. 6). As a result, almost all the motor displacement goes into the spring deflection and the force response is therefore dominated by the linear relationship between the force and the spring deflection. For the largest deformations, we recover another linear regime where the slope is nearly identical to the slope for indentation (grey shaded region in Fig.8). For this regime, we see that for a half-space (grey dashed line) the hydrodynamic repulsion increases exponentially with a slope of 1.5 in log-log plot (grey triangle), recovering the force-indentation depth relationship predicted by a Hertzian contact model (grey solid line). In this largest deformation regime, the dashed lines approach the solid lines and both the slope and intercept of the dashed lines become very close to solid lines. The near collapse of the two curves shows a recovery of static contact mechanics, although the surfaces are not in physical contact. Here, one surface acts like an indenter and the liquid between two surfaces is trapped, as discussed by others as the locked-in regime.[12, 20] In this regime, not only the $F \sim w^{1.5}$ relation for half-space indentation is recovered, but also for finite coatings, which means the corrections in contact mechanics would work for EHD case to account for effects of stratification. However, in the transition between the start-up and out-of-contact indentation there is large section of the force-central deformation curves where the dashed lines lay far above the solid lines (purple shaded region in Fig.8). In that region, the force needed to achieve a given deformation via EHD is much larger than the one needed for indentation, because both drainage and indentation occur simultaneously. As a result of the coupling between drainage and deformation the stratification approximations from contact mechanics will not be applicable for most of the drainage process. Due to these complexities, a full numerical solution is needed to characterize the full drainage process and elastic deformation during dynamic approach (except in the latest stage). The corrections based on contact mechanics can only work for one regime when fluid drainage no longer occurs.

## Conclusions

We characterized the effects of film thickness, elasticity and spring constant during fluid drainage past an incompressible stratified material. Our analysis covered all thicknesses, especially the regime of intermediate thicknesses which cannot be approximated by existing solutions for a thin or a thick film. We non-dimensionalized the governing equations to obtain three key dimensionless numbers and created contour maps to predict the deformation at contact that can be used directly to interpret experiments performed with colloidal probe microscopy or with the SFA. The key findings of our work can be summarized as follows:

- Observation of a non-linear relationship between the shift in the contact position and the thickness of the elastic coating.
- Derivation of a simple and universal analytical relationship to relate the shift in the contact position as a function of film thickness and elasticity. The relationship is valid for all experimental conditions when a stiff spring is employed.
- Demonstration that corrections for stratification derived for contact mechanics cannot be employed when the elastic deformation is caused by fluid drainage.

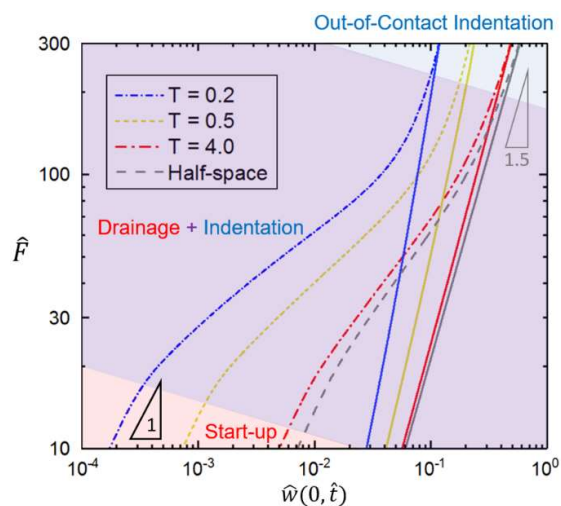

**Fig. 8:** Dashed lines: Repulsive hydrodynamic force as a function of central deformation for various thicknesses, in log-log scale. black and grey triangles indicate the slope of 1 and 1.5. Solid lines: Static indentation on soft coatings. Calculated based on previous work.[49] Grey line indicates the case of half-space, which is Hertzian contact model. All other thicknesses are indicated by corresponding colors shown as the legend. The three different background colors (red, purple, blue) indicate the three critical regimes: Start-up regime due to initiation of spring, Elastohydrodynamic drainage regime, and out-of-contact indentation regime.

## Acknowledgements

This work was supported by the National Science Foundation through NSF-CMMI 1538003. M.R.T. acknowledges support from the Johns Hopkins University Provost Undergraduate Research Award (PURA).

# Elastic Deformation of Soft Coatings Due to Lubrication Forces

Yumo Wang, Matthew R. Tan, and Joelle Frechette*
*Chemical and Biomolecular Engineering and Hopkins Extreme Materials Institute, Johns Hopkins University, Baltimore, MD 21218*

## *Supporting Information*

*1. Non-dimensionalization of the governing equations*

We non-dimensionalize our systems of equations (Equations 1-5). Naturally, the characteristic separation should be initial separation, so that the dimensionless separation is $\hat{h} = \frac{h}{h_0}$, and $\hat{h}$ start at 1 during experiments and the contact position is at $\hat{h} = 0$. We set the dimensionless radial position to be $\hat{r} = \frac{r}{r_0} = \frac{r}{\sqrt{Rh_0}}$, normalize $r$ by the initial hydrodynamic radius. If we substitute $h$ and $r$ in Equation 1 with their dimensionless variables and rearrange we get:

$$\frac{h_0}{t_0} \frac{\partial \hat{h}}{\partial \hat{t}} = \frac{1}{12\eta \hat{r}} \cdot \frac{1}{\sqrt{Rh_0}} \cdot \frac{h_0^3 p_0}{\sqrt{Rh_0}} \frac{\partial}{\partial \hat{r}}\left( \hat{r} \hat{h}^3 \frac{\partial \hat{p}}{\partial \hat{r}} \right).$$

Note that $t_0$ and $p_0$ are assumed dimensionless parameters for time and pressure. From Equation 5, we know that $Vt$ and $w$ should have the dimensionless parameter of $h_0$ (same as $h$), because they are being added or subtracted from $h$. Therefore, $t_0 = \frac{t}{\hat{t}} = \frac{h_0}{V}$. After re-arranging and cancelling terms, we have:

$$\frac{\partial \hat{h}}{\partial \hat{t}} = \frac{1}{12 \hat{r}} \cdot \frac{h_0^2 p_0}{\eta R V} \frac{\partial}{\partial \hat{r}}\left( \hat{r} \hat{h}^3 \frac{\partial \hat{p}}{\partial \hat{r}} \right).$$

We then set $p_0$ to be $\frac{\eta R V}{h_0^2}$, then:

$$\frac{\partial \hat{h}}{\partial \hat{t}} = \frac{1}{12 \hat{r}} \cdot \frac{\partial}{\partial \hat{r}}\left( \hat{r} \hat{h}^3 \frac{\partial \hat{p}}{\partial \hat{r}} \right)$$

Which is the Equation 6 in the paper. If we further put dimensionless parameters acquired so far into Equation 5:

$$k(\hat{h} - 1 + \hat{t} - \hat{w}) \cdot h_0 = \int 2\pi \hat{r} \hat{p} d\hat{r} \cdot Rh_0 \frac{\eta R V}{h_0^2}$$

Therefore:

$$\hat{F} = \int 2\pi \hat{r} \hat{p} d\hat{r} = \frac{k h_0^2}{\eta R^2 V}(\hat{h} - 1 + \hat{t} - \hat{w})$$

---

* To whom correspondence should be addressed: jfrechette@jhu.edu

Which is the Equation 10 in the paper if we denote spring parameter $K$ as $\dfrac{kh_0^2}{\eta R^2 V}$.

In equation 4, all the dimensional parameters are known, except for Hankel transform variable $\xi$. However, since $\xi r$ is dimensionless, $\hat{\xi} = \xi\sqrt{Rh_0}$, and the dimensional parameters of $\xi r$ cancel out in the Bessel function. We can render $Z$ into dimensionless term:

$$Z = \hat{\xi}\int_0^\infty \hat{r}\hat{p}J_0(\hat{\xi}\hat{r})d\hat{r} \cdot \sqrt{Rh_0} \cdot \frac{\eta RV}{h_0^2} = \hat{Z} \cdot \frac{\eta R^{1.5}V}{h_0^{1.5}}$$

From equation 3, we can see that the dimensional parameter for elastic coating thickness, $\delta$, should be $\sqrt{Rh_0}$, since $\delta$ is multiplied with $\xi$ to be the exponential order. From that, we set $T = \dfrac{\delta}{\sqrt{Rh_0}}$ to be the dimensionless coating thickness, which is also a key parameter.

If we then put all parameters into equation 2, after re-arrange and cancel terms, we have:

$$w = \hat{w} \cdot h_0 = \frac{\eta R^{1.5}V}{h_0^{1.5}} \cdot \frac{1}{E^*} \int_0^\infty \frac{2}{\hat{\xi}} X(\hat{\xi}T)\hat{Z}J_0(\hat{\xi}\hat{r})d\hat{\xi}$$

And therefore:

$$\hat{w} = \varepsilon\int_0^\infty \frac{2}{\hat{\xi}} X(\hat{\xi}T)\hat{Z}(\hat{\xi})J_0(\hat{\xi}\hat{r})d\hat{\xi},$$

with $\varepsilon = \dfrac{\eta R^{1.5}V}{h_0^{2.5}E^*}$, which is our third key dimensionless parameter.

Note that in the above analysis, the $E^*$ and $X(\xi\delta)$ terms both have the Poisson's ratio. In the current model, we set the Poisson's ratio to be 0.5 (constant), which represents incompressible materials. If the Poisson's ratio is not set to a known numerical value, the non-linear dependence of $X(\xi\delta)$ with respect to the Poisson's ratio in equation 3 makes it hard to extract the contribution of the Poison's ratio into a dimensionless parameter. This issue has been overcome in previous works [1,2] by expanding equation 3 for limiting cases (thin or thick films). For example, expand $X(\xi\delta)$ at $\xi\delta \sim 0$ (thin film) lead to $X(\xi\delta) \sim \xi\delta(1-2v)/(2(1-v)^2)$ for $v < 0.5$. Note that here the contribution of Poisson's ratio in $X(\xi\delta)$ is separated from the contribution of $\xi\delta$, and the $(1-2v)/(2(1-v)^2)$ part can be taken out from the integration in equation 2 because it's independent of $\xi$. By this method, the scaling of $v$ in $\varepsilon$ that works for all $v$ could be found, but just for thin films. Since we seek a general framework that is valid for all thicknesses and elasticity, including intermediate film thicknesses, we did not expand $X(\xi\delta)$ but set the Poisson's ratio to be a constant value of 0.5, so that the choice of elasticity parameter with respect to $v$ won't have an effect on the numerical results. Note that for Poisson's ratio < 0.5, the dimensional numerical results using our model are still CORRECT. However, a modification parameter with respect to Poisson's ratio might be needed for comparing the dimensionless results of $v = 0.5$ with $v < 0.5$ that have the same elastic parameter.

2. *Numerical Algorithm*
   2.1. *Flow chart*

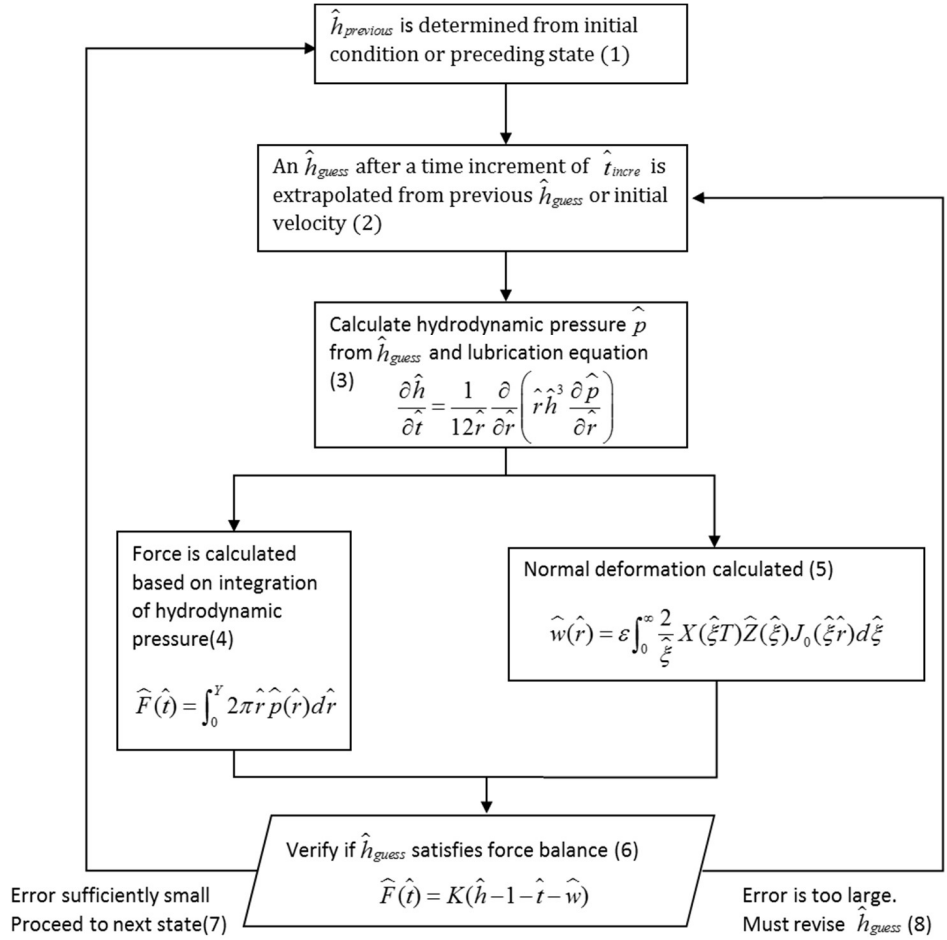

**Fig. S1**. Flow chart for the numerical algorithm employed.

2.2. *Remarks on the flow chart.*
(1) To initialize the calculation the surface at $\hat{t}=0$ was set to be undeformed and stationary and the hydrodynamic pressure is zero.
(2) The technique of backward finite difference was used to update variables such as $d\hat{h}/d\hat{t}$.
(3) Since pressure is axially symmetric, the $\frac{\partial \hat{p}}{\partial \hat{r}}(\hat{r}=0)=0$. The pressure at far side $\hat{r}=0.1\sqrt{\frac{R}{h_0}}$ ($r=0.1R$ in dimensional form) was set to be zero and pressure was neglected for $\hat{r}>0.1\sqrt{\frac{R}{h_0}}$. The vector $\hat{r}\hat{h}^3\frac{\partial \hat{p}}{\partial \hat{r}}$ could be fitted using Matlab command "spline" and "fnval".
(4) The liquid pressure for $\hat{r}>0.1\sqrt{\frac{R}{h_0}}$ is neglected. Thus, the cutoff value Y was set to be $0.1\sqrt{\frac{R}{h_0}}$.

(5) The Hankel transform variable $\xi$ was carefully meshed (1000 points) to ensure accuracy of integration. Usually the maximum value of $\xi$ taken was $\sim 10^5$. Note that some noise in integrant could be observed at very large $\xi$, if the mesh of $\hat{r}$ is not fine enough.

(6) The error in $\hat{h}_{guess}$ was transferred back into dimensional term and the tolerance criteria was set to be 0.01 nm, which is beyond the resolution of most experiments. However, a better initial $\hat{h}_{guess}$ can be gained if a smaller tolerance criteria is used, especially when the surfaces are very close.

(7) A simple method to update the new $\hat{h}_{guess}$ is to decrease the separation needed to satisfy the force balance (Step 6). ($\hat{h}^*_{guess} = 0.5(\hat{h}_{guess} + \hat{h}_{calculated})$). Here the $\hat{h}^*_{guess}$ represents the revised $\hat{h}_{guess}$ for proceeding iteration step, and $\hat{h}_{calculated}$ is the calculated separation from the force balance (Step 6) along with the hydrodynamic force from integration of pressure (Step 4). However, due to the nature of lubrication equation, at small separations, the pressure tends to be extremely sensitive to the change of separation. Therefore, a small change in $\hat{h}_{guess}$ might results in huge change in pressure, and the iteration can diverge. To improve the convergence, a weight factor ($w_f$) ranging from 0 to 1 is used in updating $\hat{h}_{guess}$, and the new $\hat{h}^*_{guess}$ is $w_f(\hat{h}_{guess} + (1-w_f)\hat{h}_{calculated})$. If $w_f$ is set to be close to 1, the change $\hat{h}_{guess}$ after each iteration is relatively small and more iterations need to be run before a satisfactory solution is reached. On the other hand, because the convergence decreases with decreasing surface separations, the weight factor need to be modified to be closer to 1 over time. This trade-off limits the efficiency of computation. In the current model, $w_f$ could be as large as 0.999 when surfaces are close.

3. *Validation of layered theory*

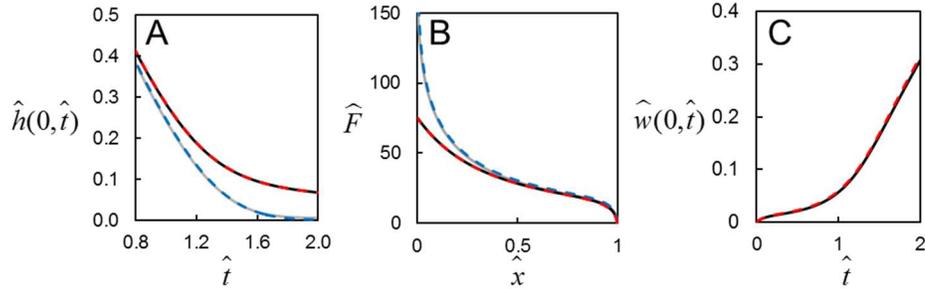

**Fig. S2.** Comparison of Elastohydrodynamic stratified theory with known limits: DSH model plotted in red dashed lines, and Reynolds' theory plotted in blue dashed lines. In all figures, Spring parameter $K$ is set to be 200, and Elasticity parameter ($\varepsilon$) for deformable surfaces are set to be 0.0026. Grey lines: T = 0.01. Black lines: T = 20. Red dashed lines: DSH theory for half space. Blue dashed lines: Reynolds' theory for rigid surfaces.

We validate our results by recovering two well-known theories: Reynolds' theory for rigid surfaces,[3] and DSH theory for soft half-space.[4] Regardless of coating material, a surface with an extremely thin coating would have negligible deformation due to constraints of a rigid substrate. In that case, it would be adequate to describe the drainage process from Taylor equation, $F = \frac{6\pi\eta R^2}{h}\frac{dh}{dt}$. In our model, if we set the thickness parameter to very small values, for example T = 0.01, we find that our results overlap with Taylor equation (see Figure S2), in which Reynolds' theory is plotted in dashed blue lines and the layered model for T = 0.01 is plotted in grey solid lines. The overlapping between two methods is found for all the central separation $\hat{h}(0,\hat{t})$, repulsive force $\hat{F}$ and central deformation $\hat{w}(0,\hat{t})$. On the other hand, in absence of substrate effects, drainage past a surface with an extremely thick compliant film will mimic that of a half-space. We take DSH model for elastic half-space and compare it with our model for the same elasticity but for a thickness parameter of T = 20, and find the force curves overlap again. In Figure S2, the red dashed lines indicate the DSH model and black solid lines indicate the new model with T = 20. Therefore we recover both the extremely thin and thick limits, by simply changing one parameter without different assumptions.

*4. Non-monotonic relations on Force vs. separation curve for varying coating thickness*

The reason for the non-monotonic dependence shown in Figure 5 is because of initialization of spring deflection in the experiments. To magnify this effect and discuss its origin, we first look at the limiting case of $K \sim \infty$, which correspond to the case of using an infinitely rigid spring compared to the compliance of surface. At the beginning of the experiments (t = 0), the motor is at rest and have a moving speed of 0, so there is no deflection, no deformation, and $h$ is kept at $h_0$. After the first time increment, the motor move $V\Delta t$ towards the other surface. In the limiting case where $K \sim \infty$, the displacement of the motor (point A in schematic Fig. S3(Left)) will be fully transmitted to the surface (point B in schematic), and the movement of point B generates drainage flow. For rigid surfaces, because of the absence of compliant coatings, there is no deformation at this step, and $h = x$. Therefore, $V = dh/dt$ = constant, so that at the first time increment, the hydrodynamic force needs to be updated from 0 to a finite value directly ( $F = \frac{6\pi\eta R^2 V}{h}$, see the "vertical wall" of rigid black line in Fig. S3(Right). For a deformable surface, however, drainage flow deforms the coating instantly and as a result the value of $dh/dt$ is no longer equal to, but smaller than $V$, because the deformation increases the separation. So the initial value of the hydrodynamic force is $F = \frac{6\pi\eta R^2}{h} \cdot \frac{dh}{dt} < \frac{6\pi\eta R^2 V}{h}$. See zoom-in of Fig. S3(Right). The thicker the coating is, the larger the difference between $dh/dt$ and $V$ because of deformation. However, at smaller central separation, for a given $h$, since a more compliant surface will have a broader interacting zone with the other surface, the repulsive force tends to be larger for the softer surface. As a result, this additional effect compensates for the initiation effect

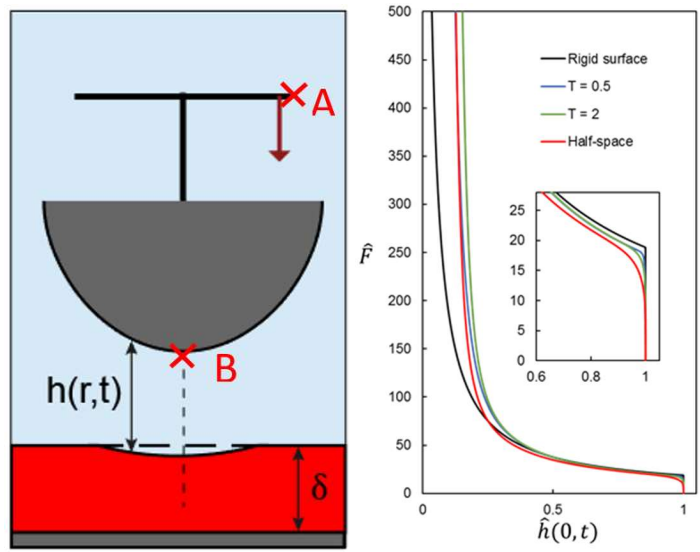

**Fig. S3.** (Left) Schematic illustration of spring initialization in the case of infinitely rigid spring. Red A and B: reference points indicated by red cross. $h(r,t)$ indicate the surface separation and $\delta$ is elastic coating thickness. (Right) Repulsive hydrodynamic force as a function of dimensionless surface separation for discussed above. So, a transition and non-monotonic effects are observed in the *f* vs *h* curve for different coating thicknesses (Fig.5A and Fig. S3). As the surfaces approach, the lubrication pressure gets much bigger, and the substrate effects is getting increasingly important, and will finally dominate over the finite initiation effects. Therefore, we would ultimately see the red line on figures below going on top of other lines, if we plot Force to large enough range.

We have run the divergence test for finer time increments and concluded that this result is not due to our artifact of the numerical method. In the case of a finite spring, for example, the $K=200$ case plotted in Figure 5A, the non-monotonic effect are much less pronounced compared to $K \sim \infty$ because the displacement of point B in schematic above now can be balanced with spring deflection, instead of directly transmitted from displacement of point A.